\documentclass[runningheads]{llncs}
\usepackage{graphicx}
\usepackage[utf8]{inputenc}
\usepackage{caption}
\usepackage{amssymb}
\usepackage{mathtools}
\usepackage{hyperref}
\usepackage{todonotes}

\usepackage{booktabs}
\usepackage{multirow}
\usepackage{adjustbox}
\usepackage{colortbl}

\usepackage{listings}

\begin{document}

\title{A Taxonomy of Anomalies in Log Data}
\titlerunning{A Taxonomy of Anomalies in Log Data}


\author{Thorsten Wittkopp \and Philipp Wiesner \and Dominik Scheinert \and Odej Kao}
\authorrunning{T. Wittkopp, et.al.}
\institute{Technische Universität Berlin\\
DOS, TU-Berlin, Germany\\
\email{\{t.wittkopp, wiesner, dominik.scheinert, o.kao\}@tu-berlin.de}}

\maketitle              

\begin{abstract}
Log data anomaly detection is a core component in the area of artificial intelligence for IT operations.
However, the large amount of existing methods makes it hard to choose the right approach for a specific system.
A better understanding of different kinds of anomalies, and which algorithms are suitable for detecting them, would support researchers and IT operators.
Although a common taxonomy for anomalies already exists, it has not yet been applied specifically to log data, pointing out the characteristics and peculiarities in this domain.

In this paper, we present a taxonomy for different kinds of log data anomalies and introduce a method for analyzing such anomalies in labeled datasets.
We applied our taxonomy to the three common benchmark datasets Thunderbird, Spirit, and BGL, and trained five state-of-the-art unsupervised anomaly detection algorithms to evaluate their performance in detecting different kinds of anomalies.
Our results show, that the most common anomaly type is also the easiest to predict.
Moreover, deep learning-based approaches outperform data mining-based approaches in all anomaly types, but especially when it comes to detecting contextual anomalies.


\end{abstract}

\section{Introduction}

The operation and maintenance of data centers and corresponding IT infrastructure are becoming increasingly difficult, due to the continuous growth of cloud computing.
To cope with this complexity, systems are using more and more levels of abstraction, leading to the creation of large multilayered systems.
However, from an IT operator perspective, these layers can even aggravate the problem by adding further technical complexity under the hood.
At the same time, unpredictable events such as downtimes can cause severe financial damage, especially in case of service level agreement~(SLA) violations~\cite{santos2017analyzing}.
The area of artificial intelligence for IT operations (AIOps) tries to manage this newly introduced complexity, by supporting cloud operators to ensure operational efficiency as well as dependability and stability~\cite{gulenko2020ai}.
A core component of AIOps systems is the detection of anomalies in monitoring data such as metrics, traces, or log data.
Especially logs are an important resource for troubleshooting, as they record events during the execution of IT applications~\cite{he2021survey}.
For this reason, a large number of methods have been proposed in the field of log data anomaly detection, mostly building on data mining~\cite{breier2015anomaly} or deep learning techniques~\cite{pang2021deep}.
While supervised methods mostly perform better in anomaly detection~\cite{zhang2019robust}, they have the drawback that the anomalies must be known at training time, which is not always the case. Furthermore, is costly and time consuming to create labeled log data, and thus unsupervised methods are of high relevance.

The wide variety of approaches to anomaly detection present IT operators with the challenge of choosing the right methods for their systems.
Although there exist some commonly used datasets for evaluating approaches such as HDFS, BGL, Thunderbird, and Spirit~\cite{oliner2007datasets,xu2009detecting}, the characteristics and properties that distinguish these datasets are often not sufficiently clarified.
Furthermore, there is no common schema on how to utilize different datasets in performance evaluations.
Hence, different anomaly detection methods perform diverse evaluations, e.g. based on time windows or individual log lines~\cite{yang2021semi,guo2021logbert,li2020swisslog,du2017deeplog}.
The evaluations of different research papers are therefore not always comparable.
It remains hard to estimate the performance of methods on new, unknown datasets based on their performance on benchmark datasets, without having more insights of the anomaly types.
We want to address this lack of understanding by making the following contributions:
\begin{itemize}
    \item We propose a taxonomy for different kinds of log data anomalies based on a well established general categorization for anomalies~\cite{chandola2009anomaly}.
    \item Using this taxonomy, we introduce a method to classify anomalies in labeled datasets and analyze the benchmark datasets BGL, Thunderbird, and Spirit.
    \item We evaluate the performance of the widely used unsupervised anomaly detection methods DeepLog, A2Log, PCA, Invariants Miner, and Isolation Forrest in detecting the different types of anomalies.
\end{itemize}

The remainder of this work is structured as follows:
\autoref{sec:towards} explains the common distinction of point and contextual anomalies and provides examples in the context of log data.
\autoref{sec:related-work} surveys the related work.
\autoref{sec:approach} introduces our taxonomy and presents a method for classifying anomalies.
\autoref{sec:evaluation} analyses three common benchmark datasets by applying our method. Furthermore, we evaluate the performance of five unsupervised anomaly detection approaches on the different types of anomalies.
\autoref{sec:conclusion} concludes the paper.

\section{Towards an Anomaly Taxonomy for Log Data}\label{sec:towards}
Labeled anomaly detection datasets contain normal samples~$\mathcal{N}$ and anomalous samples~$\mathcal{A}$. 
Each sample is described through its feature-set, which varies depending on the domain of the underlying data. 
For example, for time series data, a sample is usually described by its position in a multidimensional space and a temporal component, while in other domains, such as natural language processing, feature sets can consist of word embeddings.
A common anomaly taxonomy that is described in several works~\cite{chandola2009anomaly,sebestyen2018taxonomy,blazquez2020review} categorizes anomalies into \emph{Point Anomalies} and \emph{Contextual Anomalies}. 

\begin{figure}[htbp]
\centering
\includegraphics[width=0.5\columnwidth]{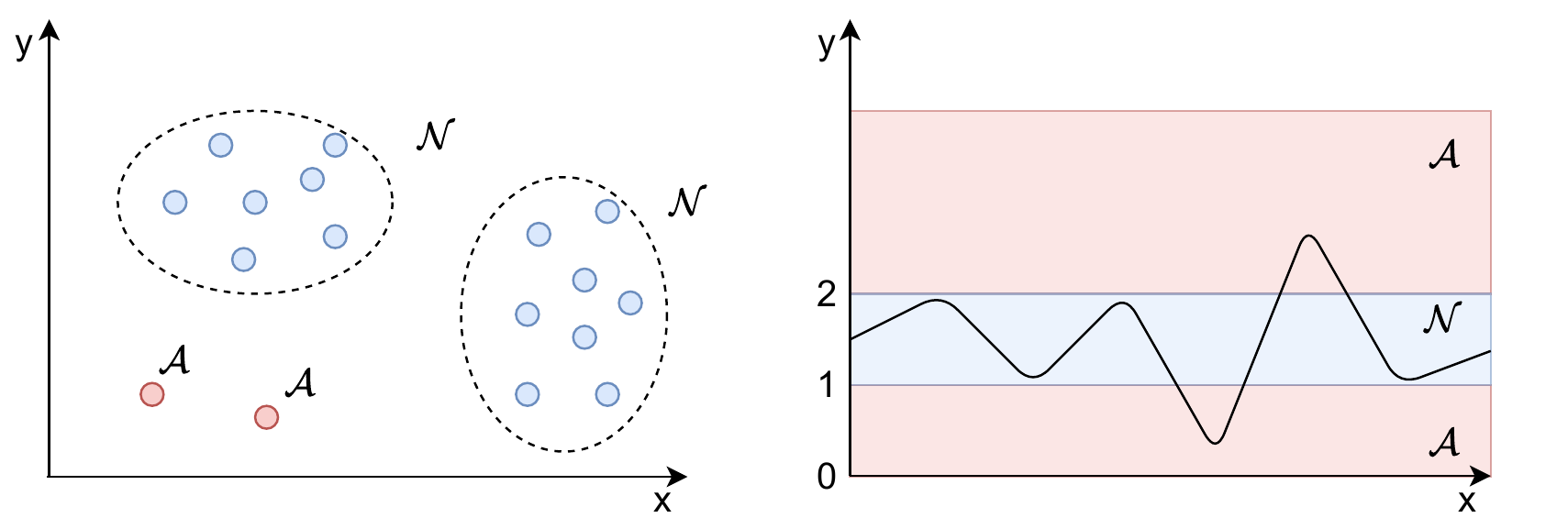}
\caption{Two examples for point anomalies. On the left side: Point anomalies in 2D space. On the right side: Point anomalies in a time series.}
\label{fig:point_anomalies_examples}
\end{figure}

\textbf{Point Anomalies.} 
A point anomaly is a single data sample that can be considered anomalous compared to the rest of the data~\cite{chandola2009anomaly}.
The values in its feature-set therefore significantly differ from the values in the feature-sets of normal samples $\mathcal{N}$.
\autoref{fig:point_anomalies_examples} illustrates two examples of point anomalies.
The first example shows two anomalies that are not located in the defined area for normal samples.
Their feature-set is the position in 2-dimensional space. 
The second example depicts point anomalies in a time series. 
As the normal feature-set is defined by $y\in[1,2]$, the anomalies are characterized by their feature-set $y\not \in[1,2]$.

\begin{figure}[htbp]
\centering
\includegraphics[width=0.4\columnwidth]{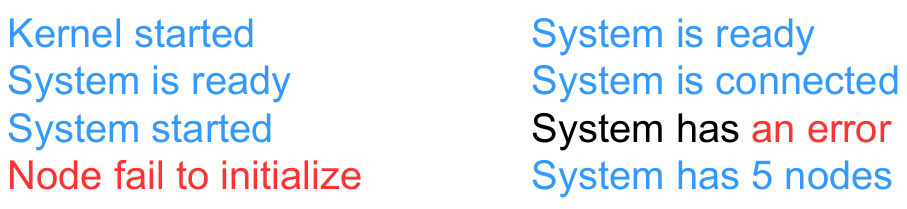}
\caption{Two examples for point anomalies in written text.}
\label{fig:point_anomalies_examples_nlp}
\end{figure}

\autoref{fig:point_anomalies_examples_nlp} depicts two point anomalies in written text. The first anomaly is trivially described through the feature-set of words. 
The anomalous sentence \emph{Node failed to initialize} has no overlapping with the feature-sets of the remaining sentences.
The second example is more fine-grained since only some words indicate an anomalous sample:
All sentences share the same or a similar prefix, only the subsequent description (\emph{ready, connected, 5 nodes, an error}) resolves the question of anomalous behavior.

\textbf{Contextual Anomalies.} Samples that are anomalous in a specific context only are called contextual anomalies~\cite{chandola2009anomaly}, and are also known as conditional anomalies~\cite{song2007conditional}.
Samples that belong to this type of anomalies can have the same feature-set (behavioral properties) as normal samples, but are still anomalous within a specific context defined by their contextual properties.



\begin{figure}[htbp]
\centering
\includegraphics[width=0.6\columnwidth]{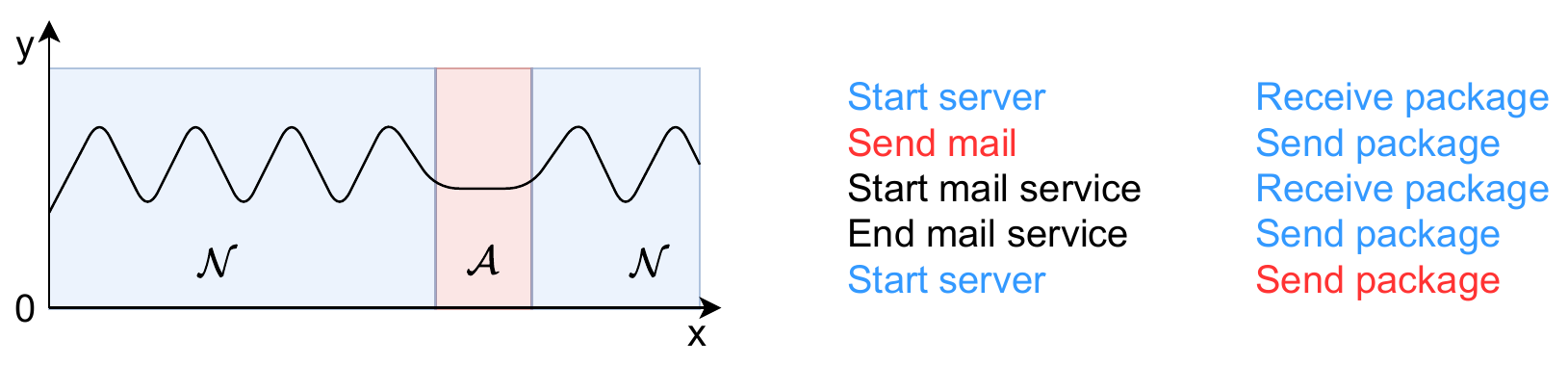}
\caption{One example of a contextual anomaly in time series and two examples for contextual anomalies in written text.}
\label{fig:context_anomalies_examples_both}
\end{figure}

\autoref{fig:context_anomalies_examples_both} illustrates this situation: 
The behavioral properties of the anomalous points (y-values / sentences) themselves are not indicating an anomaly.
However, the context of the anomalous samples defined by their contextual properties (x-values / sentence order) is different, as the normally observable strict pattern is interrupted. Furthermore,
\autoref{fig:context_anomalies_examples_both} illustrates two variants of a contextual anomaly in written text. 
In the left example, the \textit{Send mail} statement is only an anomaly because of its context, namely because
\textit{Start mail service} and \textit{End mail service} appear after the \textit{Send mail} statement. 
Since the mail service must be started before sending any mails, this ensemble of statements exemplifies a contextual anomaly. 
The second example is similar to the time series example. 
The statements \textit{Receive package} and \textit{Send package} alternate constantly. 
The anomaly is described by the fact that this alternating pattern is interrupted by a second \textit{Send package} statement.

\section{Related Work}\label{sec:related-work}

We next discuss related works with regards to defining and categorizing anomaly types, and subsequently debate concrete text-based anomaly detection methods. 

\textbf{Categorization of Anomaly Types.} An important aspect of any anomaly detection technique is the prospective nature of target anomalies.
In~\cite{chandola2009anomaly}, the authors differentiate between \emph{point anomalies}, where an individual data sample is anomalous in comparison to the rest, \emph{contextual anomalies}, where a data sample is only considered anomalous in specific contexts, and \emph{collective anomalies}, where a single data sample is only anomalous when occurring as part of a collection of related data samples, not individually.
This classification is also reused in~\cite{abs190103407}.
Similarly, the authors of~\cite{sebestyen2018taxonomy} identify the classes \emph{one-point anomaly}, \emph{contextual anomaly}, and \emph{sequential data anomaly}, and define them in the same way.
A work on outlier/anomaly detection in time series data distinguishes between \emph{point outliers} and \emph{subsequence outliers}~\cite{blazquez2020review}, which are defined as previously sketched. 
In addition, they introduce the notion of \emph{outlier time series}, where entire time series can be anomalous and are only detectable in the case of multivariate time series.
The so far highlighted types of anomalies are hence the basis for our taxonomy of anomalies in logs.
While out of our scope, log messages are further distinguishable into event log messages and state log messages~\cite{NagarajKN12} and can also be written in a distributed manner, which introduces additional challenges.

\textbf{Text-Based Anomaly Detection Methods.} In order to exemplify the diverse log anomaly types as well as illustrate the shortcomings of commonly employed methods, we make use of multiple data mining and deep learning techniques in our evaluation.
The PCA algorithm~\cite{jolliffe2005principal} is often employed for dimensionality reduction right before the actual detection procedure~\cite{he2016experience}. 
Invariant Miners~\cite{lou2010mining} retrieve structured logs using log parsing, further group log messages according to log parameter relationships, and eventually mine program invariants from the established groups in an automated fashion which are then used to perform anomaly detection on logs.
The fact that anomalies are usually few and considerably different is exploited with Isolation Forests~\cite{liu2008isolation}, an ensemble of isolation trees, where anomalies are isolated closer to the root of a tree and thus identified.
DeepLog~\cite{du2017deeplog} utilizes an LSTM and thus interprets a log as a sequence of sentences. 
It uses templates~\cite{he2017drain}, performs anomaly detection per log message, and constructs system execution workflow models for diagnosis purposes.
A2Log~\cite{wittkopp2021a2log} utilizes a self-attention neural network to obtain anomaly scores for log messages and then performs anomaly detection via a decision boundary that was set based on data augmentation of available normal training data. 

\section{Classifying Anomalies in Log Data}
\label{sec:approach}

The following chapter introduces our taxonomy for anomalies in log data. Furthermore, we present a method for classifying the anomalies in labeled datasets according to this taxonomy. 
With our classification method, system administrators are enabled to investigate their datasets and use the obtained insights to choose an appropriate anomaly detection algorithm.

\subsection{Preliminaries}
Logging is commonly employed to record the system executions by log instructions. Each instruction results in a single log message, such that the complete log is a sequence of messages $L = ( l_i : i=1,2,\ldots)$. There is a commonly used separation in \textit{meta-information} and \textit{content}. The meta-information can contain various information, for example, timestamps or severity levels. The content is free text that describes the current execution and consists of a static and a variable part.

\textbf{Tokenization.} This splits text into its segments (e.g., words, word stems, or characters).
The smallest indecomposable unit within a log content is a token. 
Consequently, each log content can be interpreted as a sequence of tokens: $s_i = (w_j: w_j \in V, j = 1,2,3,\ldots)$,
where $w$ is a token, $V$ is a set of all known tokens commonly referred to as the \textit{vocabulary}, and $j$ is the positional index of a token within the token sequence $s_i$. 

\textbf{Templates.} The tokenized log messages can be further processed into \emph{log templates}, a very common technique employed in various log anomaly detection methods~\cite{nedelkoski2020self,du2017deeplog}.
Thereby, the tokens corresponding to the static part of a log message are forming the log template $t_i$ for the i-th log message. Each unique log template is then identifiable via a log template id $x$ and referred to as $t^x$. 
All remaining tokens form the set of attributes $a_i$ for the respective log message $l_i$. 
For example the log messages: \texttt{Start mail service at node wally001} and \texttt{Start printer service at node wally005} can be described trough the template \texttt{'Start * service at node *'} with attribute sets \texttt{[mail, wally001]} and \texttt{[printer, wally005]} respectively.
Thus, each log message can be described through a log template and the attributes.

\subsection{Anomaly Taxonomy for Log Data}
In this chapter, we present our taxonomy for log data anomalies. This taxonomy relies on the categorization into \textit{Point Anomalies} and \textit{Contextual Anomalies}. 
Furthermore, we distinguish the \textit{Point Anomalies} between two types of point anomalies, namely \textit{Template Anomalies} and \textit{Attribute Anomalies}.

\begin{figure}[htbp]
\centering
\includegraphics[width=0.45\columnwidth]{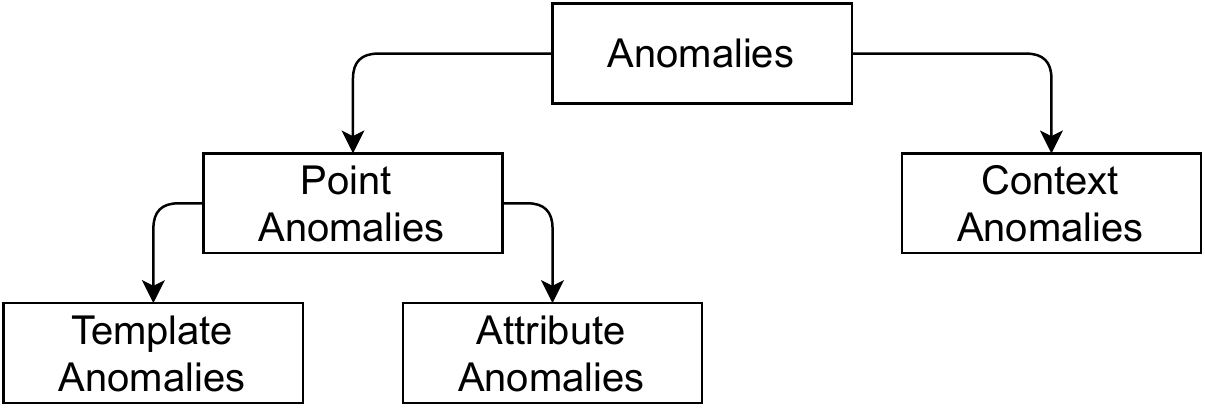}
\caption{Taxonomy for anomalies in log data.}
\label{fig:anomaly_taxonomy}
\end{figure}

\autoref{fig:anomaly_taxonomy} depicts our taxonomy. 
In the context of log data, a \textit{Point Anomaly} is an anomalous log message that is described through the log message itself. 
That is, the log message could be classified as anomalous by only investigating the respective log message and without observing its context.
The anomalous behavior of a log message is therefore described either by the corresponding template or by a specific word or number (an attribute) in the log message.
We hence define a \textit{Template Anomaly} to be characterized by the template of the respective log message. 
In contrast, an \textit{Attribute Anomaly} is described through the attributes that are extracted during the template generation process.

The second type of log data anomalies are \textit{Contextual Anomalies}.
In this case, the context, in other words the surrounding log messages, determines anomalous behavior.
The content of an individual log message is hence only relevant with respect to the log messages in its surrounding. 
In our preliminary taxonomy, we consider only single-threaded event logging scenarios for contextual anomalies.
So far, distributed logging as well as state log messages~\cite{NagarajKN12}, along with the corresponding challenges, are not yet covered.

\subsection{Anomaly Classification Method}

\begin{figure}[htbp]
\centering
\includegraphics[width=0.8\columnwidth]{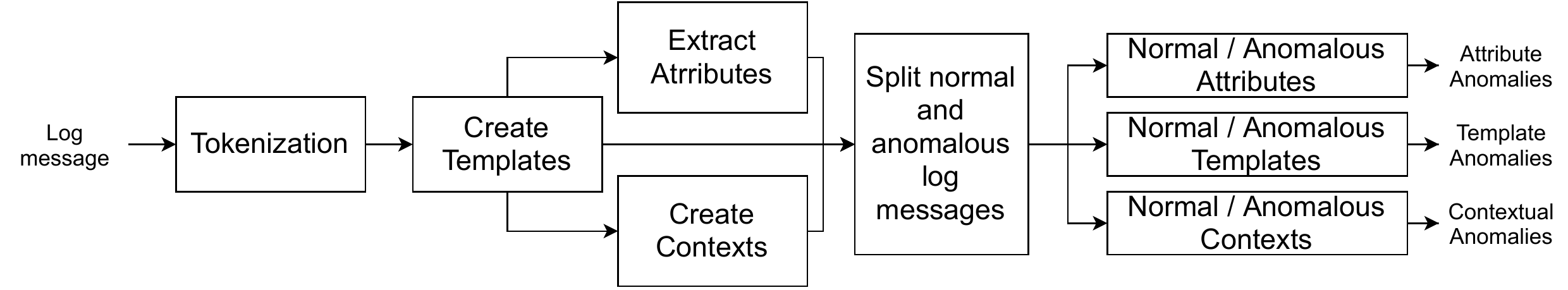}
\caption{Mining process of the different anomaly types.}
\label{fig:classification_process}
\end{figure}

The process of classifying types of anomalies in respect to our taxonomy is illustrated in~\autoref{fig:classification_process}. 
Each log message is first split into sequences of tokens in order to mine the log template. 
After all log templates are generated, we extract the attributes for each log message and calculate the context for each log line. 
The context $c_i$ for each log message $l_i$ relies on log template ids and is modelled as a set of template ids
\begin{equation}
c_i = \{ t^x_j : j = i-a,\ldots,i-1,i+1,\ldots,i+b] \},
\label{eq:context}
\end{equation}
where $a$ and $b$ are the boundaries of the context. 
For example, we calcualte the context of the 10th log message with boundaries $a=2$ and $b=1$ as $c_{10}=\{l_8,l_9, l_{11}\}$.
The template of the log message whose context is created is not considered, as described in \autoref{eq:context}. 
After deriving templates, attributes and contexts, we divide the dataset into a set of normal $\mathcal{N}$ and anomalous log messages $\mathcal{A}$ based on labels determined by experts or automated processes \cite{wittkopp2021loglab}.
Next, by utilizing these two sets and the previously calculated entities, we derive a score for each anomaly log message and each anomaly type. 
The scores represent how strong the respective anomaly type is pronounced. 
Each score is in $[0,1]$, with $1$ referring to the strongest manifestation.
\\

\textbf{Template Anomalies.}
The template anomaly $\alpha$ is calculated for each template id $x$. To get all templates for a specific template id $x$, we write $t^x(\cdot)$.
\begin{equation}
\alpha(t^x) = \frac{|t^x(\mathcal{A})|}{|t^x(\mathcal{A})|+|t^x(\mathcal{N})|}
\label{eq:template_anomaly}
\end{equation}
\vspace{3mm}

\textbf{Attribute Anomalies.} 
The attribute anomaly $\beta$ is calculated for each log message $l_i$. 
Since each log message can have multiple attributes, a score for each attribute is calculated and the attribute anomaly is then represented by the maximal score.
Here, $a_j(\cdot)$ gets all the same tokens as $a_j$ from the corresponding normal or anomalous set. 
\begin{equation}
\beta(a_i) = \max{(s: \forall a_j \in a_i. s= \frac{|a_j(\mathcal{A})|}{|a_j(\mathcal{A})|+|a_j(\mathcal{N})|})}
\label{eq:attribute_anomaly}
\end{equation}
\vspace{3mm}

\textbf{Contextual Anomalies.} 
The contextual anomaly $\gamma$ is calculated for each log message $l_i$. To get all the same contexts, for a specific context $c_i$, for each log message, we write $c_i(\cdot)$. 
\begin{equation}
\gamma(c_i) = \frac{|c_i(\mathcal{A})|}{|c_i(\mathcal{A})| + |c_i(\mathcal{N})|}
\label{eq:context_anomaly}
\end{equation}
Thus all scores can be calculated by dividing the occurrences of an event in the anomalous set by the occurrences of this event across both sets. 
As a result, $\alpha$, $\beta$, and $\gamma$ do not make an exact assignment to anomaly types but create a score that indicates how strongly it behaves to a particular anomaly type. Hence, a log line can also have several anomaly types.


\section{Evaluation}\label{sec:evaluation}

To provide an understanding on the distribution of different types of anomalies in common benchmarks according to our taxonomy, we apply our method to the Thunderbird, Spirit, and BGL datasets.
We furthermore trained five state-of-the-art unsupervised log anomaly detection methods on these datasets to evaluate their performance on predicting different types of anomalies.

The evaluation datasets were recorded at different large-scale computer systems, labeled manually by experts, and presented in \cite{oliner2007datasets}.
\autoref{table:datasets} contains the number of normal and anomalous log messages in each dataset, the amount of templates in these classes, and the number of intersecting templates.

\begin{itemize}
    \item The \emph{Thunderbird} dataset is collected from a supercomputer at Sandia National Labs (SNL) and contains more than 211 million log messages.
    
    \item The \emph{Spirit} dataset is collected from a Spirit supercomputer at SNL and contains more than 272 million log messages.
    
    \item The \emph{BGL} dataset is collected from a BlueGene/L supercomputer at Lawrence Livermore National Labs (LLNL) and contains 4\,747\,963 log messages.
\end{itemize}

\noindent From \emph{Thunderbird} and \emph{Spirit} we selected the first 5 million messages.

\begin{table}[t]
	\centering
	\caption{Dataset Statistics. Templates were generated using Drain3~\cite{he2017drain}.}
	\begin{tabular}{p{2.2cm}cc p{0.3cm} cc p{0.2cm} c}
		\toprule
        \multirow{2}{*}{Dataset}  & \multicolumn{2}{c}{Log messages} & & \multicolumn{2}{c}{Templates} && \\
        \cmidrule{2-3} \cmidrule{5-8}
                &  normal & anomalous && normal & anomalous &&  intersection \\
        \midrule
        Thunderbird   & 4\,773\,713 & 226\,287  && 969 & 17 && 3 \\
        Spirit        & 4\,235\,109 & 764\,891  && 1121 & 23 && 5 \\
        BGL           & 4\,399\,503 & 348\,460  && 802    & 58 && 10 \\
        \bottomrule
	\end{tabular}
	\label{table:datasets}
\end{table}

\subsection{Analysis of Benchmark Datasets}

\begin{figure}[h]
\centering
\includegraphics[width=0.9\columnwidth]{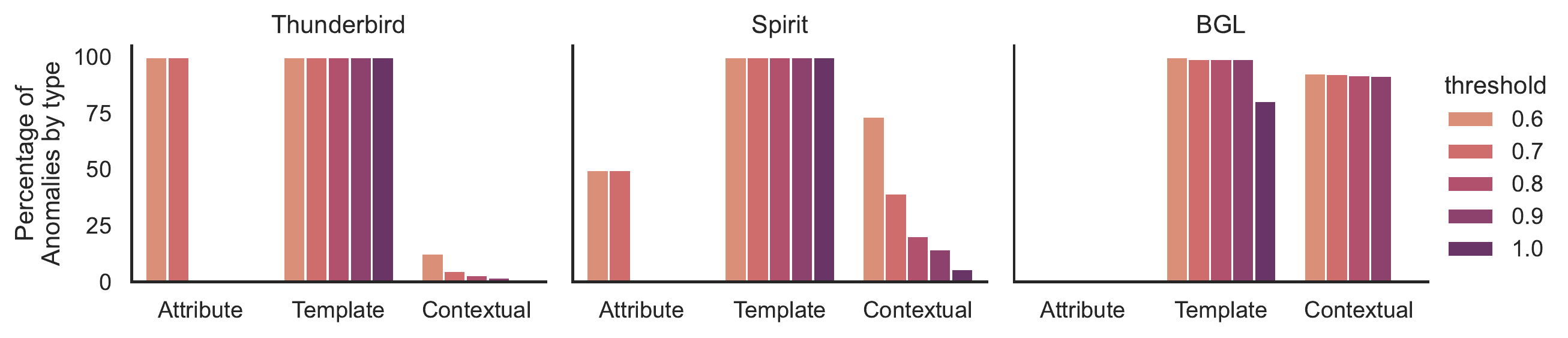}
\caption{Percentage of anomalies by type at different thresholds. Anomalies can be classified into multiple categories at the same time.}
\label{fig:approach_results}
\end{figure}

We applied our approach for classifying types of anomalies to the datasets using threshold values of 0.6, 0.7, 0.8, 0.9, and 1.0. To create the contexts we choose the following boundaries: $a=10$ and $b=0$.
\autoref{fig:approach_results} displays the results. 
We can observe that, even at threshold 1.0, more than 99\,\% of all anomalies in the Thunderbird and Spirit datasets are being classified as template anomalies. This can be explained by the fact that the intersection of normal and abnormal templates is very small.
Additionally, 226\,071 of all anomalies in Thunderbird have the same template - that is 99.9\,\%. Similarly, 99.4\,\% of all log anomalies in Spirit belong to one of the templates shown in Listing 1.
The case is very similar for BGL, although log templates are more heterogeneous in this dataset. Only at threshold 1 the amount of log messages classified as template anomalies drops to around 80\,\%.

\begin{lstlisting}[label=listing,title=Listing 1: Most anomalies in Spirit belong to these two templates (380\,271 each).,frame=tb]
kernel: hda: drive not ready for command
kernel: hda: status error: status=<:HEX:> { }
\end{lstlisting}
\vspace{0.3cm}


Until threshold 0.7 almost 100\,\% of all anomalies in Thunderbird are classifieds as attribute anomalies, meaning all template anomalies are also attribute anomalies in this case. As the "anomalous" attributes are also contained in some normal log messages, the amount of attribute anomalies drops to zero for higher thresholds.
For Spirit, we can observe that only one of the two most important log templates shown in Listing 1 contains an attribute, which explains why the number of attribute anomalies is around 50\,\%. 
At higher thresholds the number of attribute anomalies drops to zero.
For the BGL dataset no attribute anomalies were identified, not even at low thresholds.
However, more than 91\,\% of all anomalies in BGL are classified as contextual anomalies for thresholds between 0.6 and 0.9. This is significantly more than in the Thunderbird (13\,-\,2\,\%) and Spirit (63\,-\,14\,\%) datasets.

Concluding, for all three benchmark datasets algorithms that focus on detecting template anomalies are expected to perform very well.
Additionally identifying attribute anomalies may be helpful, but since most attribute anomalies are also template anomalies, the expected benefit is limited.
Approaches that identify anomalies by observing the context seem promising on datasets like BGL, but are not expected to perform well on Thunderbird and only to a certain degree on Spirit.




\subsubsection{Anomalies Outside the Taxonomy}

Our method does not guarantee that a given anomaly can be attributed to at least one of the classes in the proposed taxonomy - especially at high thresholds. 
The severity of this problem was evident to varying degrees in the datasets.
For Thunderbird, only 2 of 226\,287 messages could not be classified.
For Spirit, it was 28 out of 764\,891 for thresholds up to 0.9.
For a threshold of 1.0, we could not classify 1113 log messages, which is still only 0.15\%.
For the BGL dataset, for thresholds of 0.6, 0.7, 0.8, and 0.9, we could not classify 524, 831, 2646, and 2646 of 348460 messages, respectively. However, 68\,372 protocol messages, 19.6\% of all anomalous protocol messages, remained unlabeled at a threshold of 1.0.
Future work should either improve our classification method or describe additional types of anomalies that our proposed taxonomy does not yet cover.

\subsection{Evaluation of Unsupervised Learning Methods}

We trained five unsupervised anomaly detection algorithms to predict the different kinds of anomalies at a threshold of 0.7 in all three datasets. The goal is to identify which kinds of anomalies are easy or hard to predict and also which methods perform well on which anomalies.
In particular, we chose two deep learning approaches Deeplog~\cite{du2017deeplog} and A2Log~\cite{wittkopp2021a2log}, and three data mining approaches PCA~\cite{he2016experience}, Invarant Miners~\cite{lou2010mining}, and Isolation Forest~\cite{liu2008isolation}.
We evaluated all methods on four different train/test splits of 0.2/0.8, 0.4/0.6, 0.6/0.4, and 0.2/0.8 to test the robustness of the different methods.

\begin{figure}[bht]
\centering
\includegraphics[width=0.95\columnwidth]{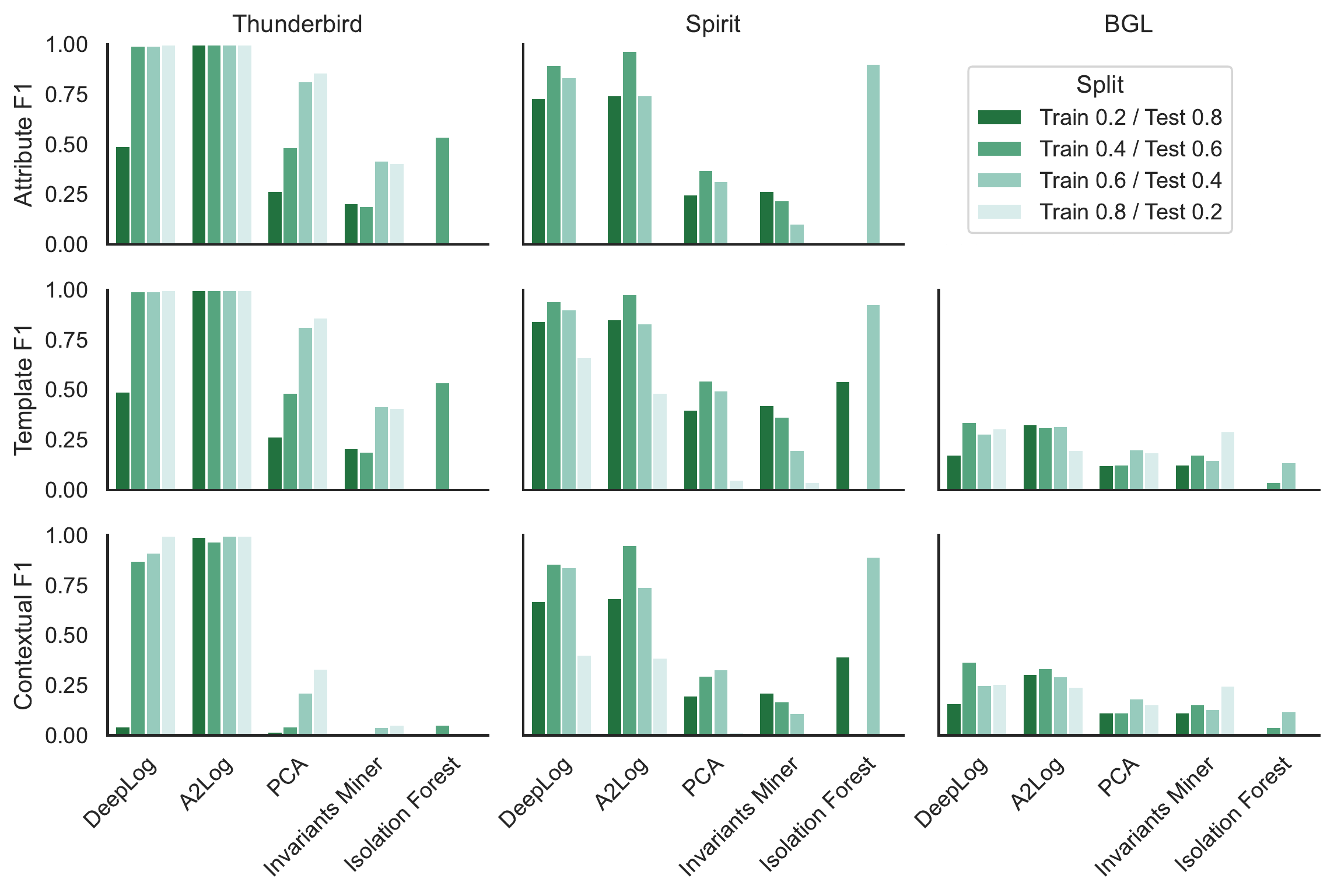}
\caption{F1 scores for predicting attribute, template, and contextual anomalies at different train/test splits at threshold 0.7.
BGL contains no attribute anomalies.}
\label{fig:ml_results}
\end{figure}

All results are depicted in \autoref{fig:ml_results}.
We can observe that the two deep learning approaches outperform the data mining approaches across all experiments.
Isolation Forrest seems to be extremely sensitive to certain "sweet spots" in train/test splits, but does not prove to be a robust method and is generally the worst-performing method. It is, hence, excluded from any further analysis in the following paragraphs.
For the Thunderbird dataset, DeepLog and A2Log manage to correctly classify almost all kinds of anomalies.
A2Log performs generally better, even when only 20\,\% of the data is available as training data.
This might be attributed to the fact that DeepLog bases its predictions on templates only, while A2Log also takes attribute information into account.
The non-deep learning methods achieve worse F1 scores on attribute and template anomalies and fail to predict most contextual anomalies.
On the Spirit dataset, the performance of all methods is generally worse.
However, the deep learning methods still achieve F1 scores of around 0.75, 0.85, and 0.7 for attribute, template, and contextual anomalies, respectively.
On BGL, all methods obtain low F1 scores for template and contextual anomalies: DeepLog around 0.27, A2Log around 0.3, PCA around 0.14, and Invariants Miner around 0.15. 
BGL does not contain any attribute anomalies.

It can be concluded, that for unsupervised methods template anomalies are the easiest to predict.
It can be suspected that attribute anomalies that are \emph{no} template anomalies are amongst the hardest to predict.
However, this is hard to show, as attribute anomalies and aemplate anomalies are highly correlated in all datasets.
Contextual anomalies were only predicted reliably by the two deep learning methods, but are generally harder to detect than template anomalies.


\section{Conclusion}\label{sec:conclusion}

In this paper, we present a taxonomy for different kinds of log data anomalies and introduce a method for applying this taxonomy on labelled datasets.
Using this method, we analyze the three common benchmark datasets Thunderbird, Spirit, and BGL.
While the vast majority of anomalies are template anomalies, BGL also contains a large number of contextual anomalies.
Attribute anomalies are highly correlated with template anomalies in all datasets.
We furthermore evaluated the ability to detect different kinds of anomalies of five state-of-the-art unsupervised anomaly detection methods: DeepLog, A2Log, PCA, Invariants Miner, and Isolation Forrest.
Our results show, that template anomalies are the easiest to predict, which explains the good performance of approaches like DeepLog.
In general, deep learning-based approaches outperform data mining-based approaches, especially when it comes to detecting contextual anomalies.

We hope that our taxonomy will enable researchers and IT Operators to better understand their datasets and help them to pick suitable anomaly detection algorithms.
Future work should investigate the log messages our approach fails to classify, potentially hinting towards further classes that are currently not present in the taxonomy.

\bibliographystyle{splncs04}
\bibliography{bib}

\begin{thebibliography}{10}
\providecommand{\url}[1]{\texttt{#1}}
\providecommand{\urlprefix}{URL }
\providecommand{\doi}[1]{https://doi.org/#1}

\bibitem{blazquez2020review}
Bl{\'a}zquez-Garc{\'\i}a, A., Conde, A., Mori, U., Lozano, J.A.: A review on
  outlier/anomaly detection in time series data. arXiv preprint
  arXiv:2002.04236  (2020)

\bibitem{breier2015anomaly}
Breier, J., Brani{\v{s}}ov{\'a}, J.: Anomaly detection from log files using
  data mining techniques. In: Information Science and Applications. Springer
  (2015)

\bibitem{abs190103407}
Chalapathy, R., Chawla, S.: Deep learning for anomaly detection: {A} survey.
  arXiv preprint arXiv:1901.03407  (2019)

\bibitem{chandola2009anomaly}
Chandola, V., Banerjee, A., Kumar, V.: Anomaly detection: A survey. ACM
  computing surveys (CSUR)  \textbf{41}(3) (2009)

\bibitem{du2017deeplog}
Du, M., Li, F., Zheng, G., Srikumar, V.: Deeplog: Anomaly detection and
  diagnosis from system logs through deep learning. In: SIGSAC (2017)

\bibitem{gulenko2020ai}
Gulenko, A., Acker, A., Kao, O., Liu, F.: Ai-governance and levels of
  automation for aiops-supported system administration. In: ICCCN. IEEE (2020)

\bibitem{guo2021logbert}
Guo, H., Yuan, S., Wu, X.: Log{BERT}: Log anomaly detection via {BERT}. In:
  International Joint Conference on Neural Networks, {IJCNN}. {IEEE} (2021)

\bibitem{he2017drain}
He, P., Zhu, J., Zheng, Z., Lyu, M.R.: Drain: An online log parsing approach
  with fixed depth tree. In: ICWS. IEEE (2017)

\bibitem{he2021survey}
He, S., He, P., Chen, Z., Yang, T., Su, Y., Lyu, M.R.: A survey on automated
  log analysis for reliability engineering. ACM Computing Surveys (CSUR)
  \textbf{54}(6) (2021)

\bibitem{he2016experience}
He, S., Zhu, J., He, P., Lyu, M.R.: Experience report: System log analysis for
  anomaly detection. In: ISSRE. IEEE (2016)

\bibitem{jolliffe2005principal}
Jolliffe, I.: Principal component analysis. Encyclopedia of statistics in
  behavioral science  (2005)

\bibitem{li2020swisslog}
Li, X., Chen, P., Jing, L., He, Z., Yu, G.: Swisslog: Robust and unified deep
  learning based log anomaly detection for diverse faults. In: ISSRE. IEEE
  (2020)

\bibitem{liu2008isolation}
Liu, F.T., Ting, K.M., Zhou, Z.H.: Isolation forest. In: 2008 eighth ieee
  international conference on data mining. IEEE (2008)

\bibitem{lou2010mining}
Lou, J.G., Fu, Q., Yang, S., Xu, Y., Li, J.: Mining invariants from console
  logs for system problem detection. In: USENIX Annual Technical Conference
  (2010)

\bibitem{NagarajKN12}
Nagaraj, K., Killian, C.E., Neville, J.: Structured comparative analysis of
  systems logs to diagnose performance problems. In: NSDI. {USENIX} Association
  (2012)

\bibitem{nedelkoski2020self}
Nedelkoski, S., Bogatinovski, J., Acker, A., Cardoso, J., Kao, O.:
  Self-supervised log parsing. In: ECML-PKDD. Springer (2020)

\bibitem{oliner2007datasets}
Oliner, A., Stearley, J.: What supercomputers say: A study of five system logs.
  In: DSN (2007)

\bibitem{pang2021deep}
Pang, G., Shen, C., Cao, L., Hengel, A.V.D.: Deep learning for anomaly
  detection: A review. ACM Computing Surveys (CSUR)  \textbf{54}(2) (2021)

\bibitem{santos2017analyzing}
Santos, G.L., Endo, P.T., Gon{\c{c}}alves, G., Rosendo, D., Gomes, D., Kelner,
  J., Sadok, D., Mahloo, M.: Analyzing the it subsystem failure impact on
  availability of cloud services. In: ISCC. IEEE (2017)

\bibitem{sebestyen2018taxonomy}
Sebestyen, G., Hangan, A., Czako, Z., Kovacs, G.: A taxonomy and platform for
  anomaly detection. In: AQTR. IEEE (2018)

\bibitem{song2007conditional}
Song, X., Wu, M., Jermaine, C.M., Ranka, S.: Conditional anomaly detection.
  {IEEE} Trans. Knowl. Data Eng.  \textbf{19}(5) (2007)

\bibitem{wittkopp2021a2log}
Wittkopp, T., Acker, A., Nedelkoski, S., Bogatinovski, J., Scheinert, D., Fan,
  W., Kao, O.: A2log: Attentive augmented log anomaly detection. In: HICSS
  (2022)

\bibitem{wittkopp2021loglab}
Wittkopp, T., Wiesner, P., Scheinert, D., Acker, A.: Loglab: Attention-based
  labeling of log data anomalies via weak supervision. In: ICSOC. Springer
  (2021)

\bibitem{xu2009detecting}
Xu, W., Huang, L., Fox, A., Patterson, D., Jordan, M.I.: Detecting large-scale
  system problems by mining console logs. In: SIGOPS. ACM (2009)

\bibitem{yang2021semi}
Yang, L., Chen, J., Wang, Z., Wang, W., Jiang, J., Dong, X., Zhang, W.:
  Semi-supervised log-based anomaly detection via probabilistic label
  estimation. In: ICSE. IEEE (2021)

\bibitem{zhang2019robust}
Zhang, X., Xu, Y., Lin, Q., Qiao, B., Zhang, H., Dang, Y., Xie, C., Yang, X.,
  Cheng, Q., Li, Z., et~al.: Robust log-based anomaly detection on unstable log
  data. In: ESEC/FSE (2019)

\end{thebibliography}
\end{document}